\documentclass{ifacconf}
\usepackage{graphicx}      
\usepackage{amssymb,fancyhdr,tabularx,amsmath,pifont}
\usepackage{url,cite,natbib}
\usepackage{float}
\usepackage{psfrag,mathptmx,arydshln}
\usepackage{latexsym,balance,multirow}
\usepackage{ifthen,colortbl}
\usepackage{color,dsfont,url}
\newcommand{\norm}[1]{\left\lVert#1\right\rVert}
\usepackage{epsfig,fancybox}
\begin{document}
\begin{frontmatter}

\title{Stability analysis of discrete-time LPV switched systems \thanksref{footnoteinfo}} 

\thanks[footnoteinfo]{This work is supported by the Brazilian agencies CNPq grants 402830/2016-4, 425800/2018-0 and FAPEMIG grant APQ-00692-17.}

\author[First]{M\'{a}rcio J. Lacerda}
\author[Second]{Cristiano M. Agulhari} 

\address[First]{ Control and Modelling Group (GCOM), Department of Electrical Engineering, Federal University of S\~ao Jo\~ao del-Rei - UFSJ, S\~ao Jo\~ao del-Rei, MG, 36307-352, Brazil. \\ E-mail: lacerda@ufsj.edu.br}
\address[Second]{Department of Electrical Engineering, Federal University of Technology of Paran\'a - UTFPR, Corn\'elio Proc\'opio, PR, 86300-000, Brazil.\\ E-mail: agulhari@utfpr.edu.br}

\begin{abstract} This paper addresses the stability problem for discrete-time switched systems under autonomous switching. Each mode of the switched system is modeled as a Linear Parameter Varying (LPV) system, the time-varying parameters can vary arbitrarily fast and are represented in a polytopic form. The Lyapunov theory is employed to get new conditions in the form of parameter-dependent LMIs. The constructed Lyapunov function takes advantage of using an augmented state vector with shifted states in its construction. In this sense, the Lyapunov function employed in this paper can be viewed as a discrete-time LPV switched Lyapunov function. Numerical experiments illustrate the efficacy of the technique in providing stability certificates.
\end{abstract}


\end{frontmatter}

\section{Introduction}

In the last decades great attention has been paid to the study of hybrid systems~\citep{GST:12}. This is due the fact that this class of systems may be used to represent several dynamics systems. The switched systems are a particular class of hybrid systems. A switched system is composed by a number of modes and each one of them can be active individually at each time. The transition between two different modes may be ruled by time, states or it can be autonomous, meaning that a transition may occur at any time~\citep{Lib:03}. 

Stability is a fundamental issue in the study of dynamical systems, including the ones with switching dynamics. In this sense, the Lyapunov theory has been successfully employed to provide stability certificates for switched systems. The Lyapunov theory allows the conditions to be written in the form of Linear Matrix Inequalities (LMIs) that can be solved via semidefinite programming~\citep{BEFB:94}. Concerning discrete-time systems with autonomous switching one may cite~\citep{DRI:02} that used a switched Lyapunov function for stability analysis and design of an output-feedback control. In~\citet{LD:06} stability conditions based on a path-dependent Lyapunov function have been exploited. The problem of stability for switched systems with time-varying delays has been investigated in~\citet{HDI:06}. In~\citet{JAPR:17} different sets of LMIs that may be used to certify stability of switched discrete-time systems are presented. Recently, a new class of switched Lyapunov functions based on the use of an augmented state vector was presented in~\cite{GL:18}.  

Even with a growing number of studies focused on stability analysis for switched systems, a small part of these studies consider the presence of uncertainties and time-varying parameters in the subsystems. Therefore, there is still great potential for the development of new less conservative and more efficient methods for switched systems. It is well known that the presence of uncertainties and time-varying parameters may affect the performance of the systems. In fact, Linear Parameter Varying (LPV) systems have been extensively studied in the last years~\citep{MS:12,Che:13a,Che:14b,Bri:15}. Thus, when analyzing switched systems it is important to consider the presence and effect of uncertainties and time-varying parameters in the stability analysis and in control design~\citep{BB:17,BB:16,BS:14}. Different approaches to the representation of uncertainties can be found in the literature, among them one can cite the polytopic uncertainties~\citep{KS:14,NR:13,RRR:12}, norm bound uncertainties \citep{ZY:15, SWX:06} and uncertainties in affine form~\citep{BS:18}. 

This paper proposes new stability conditions for discrete-time LPV switched systems under arbitrary switching. Each mode of the switched system is modeled as a LPV system in a polytopic domain. The time-varying parameters can vary arbitrarily fast and there is no information about their rates of variation. Stability will be guaranteed by means of a Lyapunov function composed by an augmented state vector. This class of function allows to introduce the switched dynamics of the system and the LPV feature in the Lyapunov function. In this sense, the Lyapunov function employed in this paper can be viewed as a discrete-time LPV switched Lyapunov function. This methodology is based upon the methods presented in~\citet{GL:18}, concerned with stability problem for precisely known switched systems, and in~\citet{LG:20}, where the stability and stabilizability problem have been considered. The use of structured Lyapunov functions with non-monotonic terms was explored to deal with the stability problem for uncertain systems in~\cite{LS:17}, moreover, stability and performance for uncertain systems were investigated using an augmented state-vector in the Lyapunov function~\citep{PLA:18,PLL:19}. The main objective of this paper is to propose less conservative conditions to guarantee stability of discrete-time switched LPV systems. The key feature in this paper is the use of shifted states, for instance $x\left ( k+1 \right )=A_{\sigma\left(k\right)}\left(\alpha_k\right) x\left ( k \right )$, implying that $x\left ( k+2 \right )=A_{\sigma\left(k+1\right)}\left(\alpha_{k+1}\right) x\left ( k+1 \right )$ or simply $x\left ( k+2 \right )=A_{\sigma\left(k+1\right)}\left(\alpha_{k+1}\right) A_{\sigma\left(k\right)}\left(\alpha_k\right) x\left ( k \right )$. Note that both the time varying parameter  $\alpha_k$ and the switching rule $\sigma\left(k\right)$ are evaluated in different instants. This fact have been investigated in~\cite{DB:01} for LPV systems, in~\cite{DRI:02} for switched systems and in~\cite{HDI:06b} for switched LPV systems considering only two different instants. The approach addressed in this paper admits the use of a generic number of shifted states and consequently a generic number of instants in the swithched rule and also in the LPV parameter. Numerical examples borrowed from the literature are employed to illustrate the advantages of the proposed technique when compared to existing approaches.

This paper is organized as follows. Preliminary results are presented in Section~\ref{rpr}, Section~\ref{mre} details the main contributions of the paper. The performance of the method is illustrated via numerical experiments in Section~\ref{nex}, while Section~\ref{con} concludes the paper.



\section{Preliminaries}

\label{rpr}
\subsection{System description}

Consider the following switched discrete-time LPV system
\begin{equation}
\label{eq1}
    x\left ( k+1 \right )=A_{\sigma\left(k\right)}\left(\alpha_k\right) x\left ( k \right )
\end{equation}
where $x \in \mathbb{R}^n$ is the state vector, $A_{\sigma\left(k\right)}\left(\alpha_k\right) \in \mathbb{R}^{n \times n}$ is the dynamic matrix, $\sigma(k)$, belongs to a finite set $\mathcal{P}$ that denotes the switching rule $\mathcal{P} = \left\{1,\ldots, m\right\}$, $\alpha_k$ is the time-varying parameter that belongs to a polytopic domain parameterized in terms of a vector of time-varying parameters. Although each mode could be subject to a different time-varying parameter, to simplify the developments, let us consider that all the modes present the same number of vertices and are affected by the same time-varying parameter $\alpha_k$.

For a specific mode $\sigma(k)$ it is possible to write
\begin{equation*}
\label{politopo} 
    A_{\sigma(k)}(\alpha_k) = \sum_{i=1}^V \alpha_{k,i} A_{\sigma(k),i}, \quad \alpha_k \in \Lambda_V
\end{equation*}
where $A_{\sigma(k),i}$, $i=1,\ldots, V$, are the vertices of the polytope  and $\Lambda_V$ is the unit simplex given by
\begin{equation*}
    {\Lambda_{V}= \left \{ \alpha_k \in \mathbb{R}^{V} : \sum_{i=1}^{V} \alpha_{k,i} = 1 ; \alpha_{k,i}\geq 0, i=1,\ldots,V \right \}}.
\end{equation*}

Only one mode of the matrix $A_{\sigma(k)}$ is active at a time. The indicator function will be used to describe such a behavior. Consider
$
 {   \xi \left ( k \right ) = \left [ \xi_{1}\left ( k \right ),\ldots , \xi_{m}\left ( k \right ) \right ]^{T}}
$
\begin{equation*}
\begin{aligned}
    \xi_{i}\left(k \right ) = \left\{\begin{matrix}1, \quad &
    \quad & \mbox{if}~\sigma(k) = i \\ 
    0,& \quad & \mbox{otherwise.}\end{matrix}\right.
\end{aligned}
\end{equation*}
 In this way, system~\eqref{eq1} can be written as
\begin{equation}
\label{eq2}
     x\left(k+1 \right ) = A\left(\xi\left(k \right ),\alpha_k \right)x\left(k \right ).
\end{equation}

\subsection{Stability analysis}
Stability of system~\eqref{eq1} can be certified by the existence of a radially unbounded Lyapunov function $V(k,x(k))$ satisfying the following criteria~\citep{Vid:93}     
\begin{equation}
\label{eq3}
V(k,0)= 0, \quad V(k,x(k)) > 0, \quad  \forall x(k) \neq 0,
\end{equation}
\begin{equation}
\label{eq4}
    \Delta{V}(k,x(k))  < 0, \quad \forall x(k) \neq 0,
\end{equation}
where $\Delta{V}(k,x(k))= V(k+1,x(k+1)) - V(k,x(k))$. The Lyapunov function satisfies
	\begin{equation}
	\label{eqs}
	\beta_1 \norm{x(k)}^2 \leq V(k,x(k)) \leq \beta_2 \norm{x(k)}^2
	\end{equation}
	for all $x(k) \in \mathbb{R}^n$ and $k \geq 0$ with $\beta_1$ and $\beta_2$ positive scalars. Moreover, $\Delta{V}(k,x(k))< -\beta_3 \norm{x(k)}^2$, where $\beta_3$ is a sufficiently small positive scalar. If such a Lyapunov function exists, then system~\eqref{eq2} is GUAS (Globally Uniformly Asymptotically Stable).

In~\cite{HDI:06b} a set of conditions for robust stability analysis of switched systems is proposed, where each switching mode is described by a polytopic domain represented by a vector of time-varying parameters. The following lemma presents the main result of such paper.

\begin{lem}
\label{lema2}
If there exist symmetric positive definite matrices $S_i(\alpha_k), S_j(\alpha_{k+1})$ and matrices $G_i(\alpha_k)$ of appropriate dimensions such that
\begin{equation}
\label{eq8}
\begin{bmatrix}
G_i(\alpha_k) + G_i(\alpha_k)^T - S_i(\alpha_k) & \star \\
A_i(\alpha_k)G_i(\alpha_k) & S_j(\alpha_{k+1})
\end{bmatrix} > 0
\end{equation}
$\forall \alpha_k \in \Lambda_V, \alpha_{k+1} \in \Lambda_V$, $i \in \mathcal{P}, j \in \mathcal{P}$, then system~\eqref{eq2} is GUAS.
\end{lem}

\begin{pf}
Since
\[
G_i(\alpha_k)^TS_i(\alpha_k)^{-1}G_i(\alpha_k) \geq G_i(\alpha_k) + G_i(\alpha_k)^T - S_i(\alpha_k),
\]
condition~\eqref{eq8} implies
\[
\begin{bmatrix}
G_i(\alpha_k)^{T}S_i(\alpha_k)^{-1}G_i(\alpha_k) & \star \\
A_i(\alpha_k)G_i(\alpha_k) & S_j(\alpha_{k+1})
\end{bmatrix} > 0.
\]
Pre- and post-multiplying the latter condition respectively by $\mbox{diag}(G_i(\alpha_k)^{-T}, S^{-1}_j(\alpha_{k+1}))$ and its transpose, and setting $S^{-1}_i(\alpha_k) = P_i(\alpha_k)$, results in
\[
\begin{bmatrix}
P_i(\alpha_k) & \star \\
P_j(\alpha_{k+1})A_i(\alpha_k) & P_j(\alpha_{k+1})
\end{bmatrix} > 0.
\]
The application of a Schur complement~\citep{BEFB:94} yields
\begin{equation}
\label{eq_lyap1}
    A_i(\alpha_k)^TP_j(\alpha_{k+1})A_i(\alpha_k) - P_i(\alpha_k) < 0.
\end{equation}
Multiplying~\eqref{eq_lyap1} by $\xi_i(k)^2$, $i = 1, \ldots, m$, and summing up gives
\begin{equation}
\label{eq_lyap2}
    A(\xi(k),\alpha_k)^T P_j(\alpha_{k+1}) A(\xi(k),\alpha_k) - P(\xi(k),\alpha_k) < 0.
\end{equation}
Multiplying~\eqref{eq_lyap2} by $\xi_j(k+1)$, $j = 1, \ldots, m$, and summing up results in
\begin{equation}
\label{eq_lyap3}
    A(\xi(k),\alpha_k)^T P(\xi(k+1),\alpha_{k+1}) A(\xi(k),\alpha_k) - P(\xi(k),\alpha_k) < 0,
\end{equation}
which is equivalent to condition~\eqref{eq4} with 
\[
V(k,x(k)) = x(k)^T P(\xi(k),\alpha_k) x(k).
\]
Since $P(\xi(k),\alpha_k) > 0$, the Lyapunov function $V(k,x(k))$ is also positive definite, concluding the proof.
\end{pf}

The condition presented in Lemma~\ref{lema2} depends on additional slack variables that, although reducing the conservativeness, increase the computational cost to solve the problem. In the following section, an alternative way to assess the stability of discrete-time LPV switched systems is proposed, based on the utilization of augmented Lyapunov functions.


\section{Main results}
\label{mre}

This paper employs a class of structured Lyapunov functions to provide stability certificates for switched discrete-time LPV systems. This class of Lyapunov functions introduces the dynamics of the system in its construction. To better illustrate our approach, firstly we will provide a formulation based on a particular case. 
\begin{lem}
\label{lema3}
If there exist symmetric matrices $P_{1} \in \mathbb{R}^{n\times n}$ and $P_{2} \in \mathbb{R}^{n\times n}$ such that
\begin{equation}
\label{eq12}
     P_{1} + A_{i}\left(\alpha_k \right )^{T}P_{2}A_{i}\left(\alpha_k \right )>0
\end{equation}

\begin{multline}
\label{eq13}
    A_{i}\left(\alpha_k \right )^{T}P_{1}A_{i}\left(\alpha_k \right ) +
    A_{i}\left(\alpha_k \right )^{T}A_{j}\left(\alpha_{k+1} \right )^{T}P_{2}A_{j}\left(\alpha_{k+1} \right )A_{i}\left(\alpha_k \right )\\ - \left(P_{1}+A_{i}\left(\alpha_k \right)^{T}P_{2}A_{i}\left(\alpha_k \right ) \right ) < 0
\end{multline}
$\forall~ \alpha_k \in \Lambda_{V}$, $\alpha_{k+1} \in \Lambda_{V}$, $i \in \mathcal{P}$, $j \in  \mathcal{P}$, then system~\eqref{eq2} is GUAS. 
\end{lem}

\begin{pf}
By multiplying~\eqref{eq13} by $\xi_i(k)^2$, $i=1,\ldots,m$, and summing up one has
\begin{multline}
\label{eq14}
A(\xi(k),\alpha_k)^TP_1A(\xi(k),\alpha_k)^T\\
+A(\xi(k),\alpha_k)^TA_j(\alpha_{k+1})^TP_2A_j(\alpha_{k+1})A(\xi(k),\alpha_k) \\
-\left(P_{1}+A(\xi(k),\alpha_k)^{T}P_{2}A(\xi(k),\alpha_k) \right ) < 0
\end{multline}
Multiplying~\eqref{eq14} by $\xi_j(k+1)^2$, $j=1,\ldots,m$, and summing up one has 
\begin{multline}
\label{eq15}
A(\xi(k),\alpha_k)^TP_1A(\xi(k),\alpha_k)^T
+\Upsilon^TP_2\Upsilon \\
-\left(P_{1}+A(\xi(k),\alpha_k)^{T}P_{2}A(\xi(k),\alpha_k) \right ) < 0
\end{multline}
with
\begin{equation*}
      \Upsilon = A\left(\xi\left(k+1 \right ),\alpha_{k+1} \right )A\left(\xi\left(k \right ),\alpha_k \right ). 
\end{equation*}
Pre- and post-multiplying~\eqref{eq15} by $x(k)^T$ and $x(k)$ respectively and considering the dynamics of the system, i.e., $ x\left(k+1 \right ) = A\left(\xi\left(k \right ),\alpha_k \right)x\left(k \right )$ and  $x\left(k+2 \right ) = A\left(\xi\left(k+1 \right ),\alpha_{k+1} \right)x\left(k+1 \right )$ yields
\[
V(x(k+1))-V(x(k))<0
\]
with $V(x(k))=x(k)^T \left(P_1+A(\xi(k),\alpha_k)^TP_2A(\xi(k),\alpha_k)\right)x(k)$. Note that, by multiplying~\eqref{eq12} by $\xi_i(k)^2$, $i=1,\ldots,m$, and summing up one has
\begin{equation}
    P_1+A(\xi(k),\alpha_k)^TP_2A(\xi(k),\alpha_k)>0,
\end{equation}
ensuring that the Lyapunov function $V(x(k))$ is positive definite. Moreover, one may choose

\begin{align*}
\beta_1&= \min_{i \in \mathcal{P}, \alpha_k \in \Lambda_V} \lambda_{\min} \left( P_{1} + A_{i}\left(\alpha_k \right )^{T}P_{2}A_{i}\left(\alpha_k \right ) \right)  \\    
\beta_2&= \max_{i \in \mathcal{P}, \alpha_k \in \Lambda_V} \lambda_{\max} \left( P_{1} + A_{i}\left(\alpha_k \right )^{T}P_{2}A_{i}\left(\alpha_k \right ) \right)    
\end{align*}
to guarantee that~\eqref{eqs} is satisfied and
\begin{multline*}
    \beta_3= \min_{i,j \in \mathcal{P}, \alpha_k,\alpha_{k+1} \in \Lambda_V} \lambda_{\min} \left( A_{i}\left(\alpha_k \right )^{T}P_{1}A_{i}\left(\alpha_k \right ) \right.  \\
    +
    A_{i}\left(\alpha \right )^{T}A_{j}\left(\alpha_{k+1} \right )^{T}P_{2}A_{j}\left(\alpha_{k+1} \right )A_{i}\left(\alpha_k \right )\\ \left. - P_{1}-A_{i}\left(\alpha_k \right)^{T}P_{2}A_{i}\left(\alpha_k \right ) \right)
\end{multline*}
to ensure $\Delta{V}(k,x(k))< -\beta_3 \norm{x(k)}^2$, concluding the proof.
\end{pf}

\begin{rem}
Note that even considering constant matrices $P_1$ and $P_2$, the Lyapunov function 
\[
V(x(k))=x(k)^T \left(P_1+A(\xi(k),\alpha_k)^TP_2A(\xi(k),\alpha_k)\right)x(k)
\]
depends upon the switching modes $\xi(k)$ and the LPV parameter~$\alpha_k$. Moreover, there is no sign constraints imposed to the symmetric matrices $P_1$ and $P_2$ individually. It is also simple to verify that, considering $P_2=0$ in Lemma~\ref{lema3}, allow us to recover the results presented in~\eqref{eq_lyap1}.
\end{rem}

In what follows the result presented in Lemma~\ref{lema3} will be extended to the more general case making use of $N$ symmetric matrices $P_i$. Before introducing the main results let us define some notation. Consider
\begin{align*}
    \Phi_0 &= I \\
    \Phi_1 &= A_{i_1}(\alpha_k) \\
    \Phi_2 &= A_{i_2}(\alpha_{k+1})A_{i_1}(\alpha_{k}) \\
    \Phi_R &= A_{i_R}(\alpha_{k+R-1})A_{i_{R-1}}(\alpha_{k+R-2})\cdots A_{i_1}(\alpha_{k}).
\end{align*}
Moreover, a multi-simplex domain composed by the cartesian product of $N$ different simplex sets, each of them with $V$ vertices, is denoted by $\Lambda_V^N$. In other words
\[
\Lambda_V^N= \underbrace{  \Lambda_V \times \Lambda_V \times \ldots \times \Lambda_V}_{N-times}.
\]
In the same way
\[
\mathcal{P}^N= \underbrace{  \mathcal{P} \times \mathcal{P} \times \ldots \times \mathcal{P}}_{N-times}
\]
is the cartesian product of  finite sets $\mathcal{P}$.


\begin{thm}
\label{teo1}
If there exist symmetric matrices $P_{i} \in \mathbb{R}^{n\times n}$, $i=1,\ldots,N$, such that
\begin{equation}
\label{eq18}
    \sum_{j=0}^{N-1} \Phi_j^{T}P_{j+1}\Phi_j > 0 \\
\end{equation}
\begin{equation*}
        \forall \left(i_{1},\ldots,i_{N-1}\right) \in \mathcal{P}^{N-1},~ \forall \left(\alpha_{k},\ldots,\alpha_{k+N-2}\right) \in \Lambda_V^{N-1}
\end{equation*}

\begin{equation}
\label{eq19}
\sum_{z=1}^{N} \Phi_z^{T}P_{z}\Phi_z -    \sum_{j=0}^{N-1} \Phi_j^{T}P_{j+1}\Phi_j < 0 \\
\end{equation}
\begin{equation*}
        \forall \left(i_{1},\ldots,i_{N}\right) \in \mathcal{P}^{N},~ \forall \left(\alpha_{k},\ldots,\alpha_{k+N-1}\right) \in \Lambda_V^{N}
\end{equation*}
then system~\eqref{eq2} is GUAS. 
\end{thm}

\begin{pf}
By multiplying~\eqref{eq19} successively by $\xi_{i_j}(k+j-1)^2$, $j=1,\ldots N$, $i_j \in \mathcal{P}$ and summing up yields
\begin{equation}
\label{eqp1}  
A(\xi(k),\alpha_k)^TM_{k+1}A(\xi(k),\alpha_k)-M_{k}<0
\end{equation}
with
\begin{multline*}
M_{k}=P_1+A(\xi(k),\alpha_k)^{T}P_{2}A(\xi(k),\alpha_k) \\ + \Psi_2(k)^{T}P_{3}\Psi_2(k) + \ldots + \Psi_{N-1}(k)^{T}P_{N}\Psi_{N-1}(k)
\end{multline*}
where
\begin{align*}
    \Psi_2(k) &= A(\xi(k+1),\alpha_{k+1})A(\xi(k),\alpha_{k}) \\
    \Psi_3(k) &= A(\xi(k+2),\alpha_{k+2})A(\xi(k+1),\alpha_{k+1})A(\xi(k),\alpha_{k}) \\
    \vdots
    \\
    \Psi_N(k) &= A(\xi(k+N-1),\alpha_{k+N-1}) \times \\ &A(\xi(k+N-2),\alpha_{k+N-2}) \cdots  A(\xi(k),\alpha_{k}). 
\end{align*}
Pre- and post multiplying~\eqref{eqp1} by $x(k)^T$ and $x(k)$ respectively, and considering the dynamics of the system $ x\left(k+1 \right ) = A\left(\xi\left(k \right ),\alpha_k \right)x\left(k \right )$ one can write
\[
x(k+1)^TM_{k+1}x(k+1)-x(k)^TM_kx(k)<0
\]
that is equivalent to $V(x(k+1))-V(x(k))<0$ with $V(x(k))=x(k)^TM_kx(k)$. Note that~\eqref{eq18} guarantees that $M_k$ is positive definite. The same procedure adopted in Lemma~\ref{lema3} can be used to choose the scalars $\beta_1$, $\beta_2$ and $\beta_3$, concluding the proof.
\end{pf}

\begin{rem}
The number of scalar decision variables ($N_V$) spent by Theorem~\ref{teo1} can be computed as
\[
N_V= \frac{Nn(n+1)}{2}
\]
where $n$ is the number of states and $N$ is the number of employed matrices $P_i$. The number of LMI rows ($N_R$) can be computed as
\[
N_R=nm^{N-1} \left(\frac{(V+1)!}{2!(V-1)!}\right)^{N-1}+ nm^{N} \left(\frac{(V+1)!}{2!(V-1)!}\right)^{N}.
\]
\end{rem}

If the system is precisely known, the conditions presented in Theorem~\ref{teo1} recover the results presented in~\cite[Theorem~5]{GL:18}. The conditions presented in Theorem~\ref{teo1} can be easily adapted to consider time-invariant uncertainties. For this end, it suffices to consider $\alpha_{k+\theta}=\alpha$, for all values of~$\theta$.

To reduce the conservativness of Theorem~\ref{teo1} it is possible to introduce parameter dependent matrices $P_i(\alpha_k)$.

\begin{cor}
\label{cor1}
If there exist symmetric matrices $P_{i}(\alpha_k) \in \mathbb{R}^{n\times n}$, $i=1,\ldots,N$, such that
\begin{equation}
    \sum_{j=0}^{N-1} \Phi_j^{T}P_{j+1}(\alpha_k)\Phi_j > 0 \\
\end{equation}
\begin{equation*}
        \forall \left(i_{1},\ldots,i_{N-1}\right) \in \mathcal{P}^{N-1},~ \forall \left(\alpha_{k},\ldots,\alpha_{k+N-2}\right) \in \Lambda_V^{N-1}
\end{equation*}

\begin{equation}
\sum_{z=1}^{N} \Phi_z^{T}P_{z}(\alpha_{k+1})\Phi_z -    \sum_{j=0}^{N-1} \Phi_j^{T}P_{j+1}(\alpha_k)\Phi_j < 0 \\
\end{equation}
\begin{equation*}
        \forall \left(i_{1},\ldots,i_{N}\right) \in \mathcal{P}^{N},~ \forall \left(\alpha_{k},\ldots,\alpha_{k+N-1}\right) \in \Lambda_V^{N}
\end{equation*}
then system~\eqref{eq2} is GUAS. 
\end{cor}

\begin{pf}
The proof follows the same steps presented in the proof of Theorem~\ref{teo1}.
\end{pf}

All the conditions presented until this point are in the form of parameter-dependent LMIs that depends upon $\alpha_{k+N}$. In order to get a finite set of LMIs, in terms of the vertices of each switched mode, the ROLMIP package was employed~\citep{AOP:19}. To illustrate the process employed to write the LMIs, the conditions of Lemma~\ref{lema3} will be presented in a finite form. 

\begin{lem}\label{lem_finite}
If there exist symmetric matrices $P_1 \in \mathbb{R}^{n \times n}$ and $P_2 \in \mathbb{R}^{n \times n}$ such that, $\forall i \in \mathcal{P}, j \in \mathcal{P}$, one has
\begin{equation}\label{eq_finite_1}
    P_1 + A_{i,\ell}^T P_2 A_{i,\ell} > 0, \quad  \ell = 1, \ldots, V,
\end{equation}
\begin{multline}\label{eq_finite_2}
    2P_1 + A_{i,\ell}^T P_2 A_{i,q} + A_{i,q}^T P_2 A_{i,\ell} > 0, \\  \ell = 1, \ldots, V-1, \quad  q = \ell+1, \ldots, V,
\end{multline}
\begin{multline}\label{eq_finite_a1b1}
A_{i,\ell}^T P_1 A_{i,\ell} + A_{i,\ell}^T A_{j,r}^T P_2 A_{j,r} A_{i,\ell} - (P_1 + A_{i,\ell}^T P_2 A_{i,\ell}) < 0,  \\
 \ell = 1, \ldots, V, \quad r = 1, \ldots, V
\end{multline}
\begin{multline}\label{eq_finite_a1b1b2}
2A_{i,\ell}^T P_1 A_{i,\ell} + 
A_{i,\ell}^T A_{j,r}^T P_2 A_{j,p} A_{i,\ell} +
A_{i,\ell}^T A_{j,p}^T P_2 A_{j,r} A_{i,\ell} \\ 
-(2 P_1 + 2 A_{i,\ell}^T P_2 A_{i,\ell}) < 0, \\
\ell = 1, \ldots, V, \quad r = 1, \ldots, V-1, \quad p = r+1, \ldots, V,
\end{multline}
\begin{multline}\label{eq_finite_a1a2b1}
A_{i,\ell}^T P_1 A_{i,q} + A_{i,q}^T P_1 A_{i,\ell} +
A_{i,\ell}^T A_{j,r}^T P_2 A_{j,r} A_{i,q} +
A_{i,q}^T A_{j,r}^T P_2 A_{j,r} A_{i,\ell} \\
- (2 P_1 + A_{i,\ell}^T P_2 A_{i,q} + A_{i,q}^T P_2 A_{i,\ell}) < 0, \\
\ell = 1, \ldots, V-1, \quad q = \ell+1, \ldots, V,
\quad r = 1, \ldots, V
\end{multline}
\begin{multline}\label{eq_finite_a1a2b1b2}
2A_{i,\ell}^T P_1 A_{i,q} + 2A_{i,q}^T P_1 A_{i,\ell} +
A_{i,\ell}^T A_{j,r}^T P_2 A_{j,p} A_{i,q} \\ +
A_{i,q}^T A_{j,r}^T P_2 A_{j,p} A_{i,\ell} +
A_{i,\ell}^T A_{j,p}^T P_2 A_{j,r} A_{i,q} +
A_{i,q}^T A_{j,p}^T P_2 A_{j,r} A_{i,\ell} \\ -
(4 P_1 + 2A_{i,\ell}^T P_2 A_{i,q} + 2A_{i,q}^T P_2 A_{i,\ell}) < 0, \\
\ell = 1, \ldots, V-1, \quad q = \ell+1, \ldots, V,
\\ r = 1, \ldots, V-1, \quad p = r+1, \ldots, V,
\end{multline}
then system~\eqref{eq2} is GUAS.
\end{lem}

\begin{pf}
Multiplying~\eqref{eq_finite_1} by $\alpha_{k,\ell}^2$ and~\eqref{eq_finite_2} by $\alpha_{k,\ell}\alpha_{k,r}$, adding both results and summing up variables $\ell$ and $r$ in the respective domains yields~\eqref{eq12}.
Multiplying~\eqref{eq_finite_a1b1} by $\alpha_{k,\ell}^2\alpha_{k+1,r}^2$, \eqref{eq_finite_a1b1b2} by 
$\alpha_{k,\ell}^2\alpha_{k+1,r}\alpha_{k+1,p}$, \eqref{eq_finite_a1a2b1} by $\alpha_{k,\ell}\alpha_{k,q}\alpha_{k+1,r}^2$ and~\eqref{eq_finite_a1a2b1b2} by
$\alpha_{k,\ell}\alpha_{k,q}\alpha_{k+1,r}\alpha_{k+1,p}$,
adding all the results and summing up variables $\ell, r, p$ and $q$ in the respective domains yields~\eqref{eq13}, concluding the proof.
\end{pf}

\begin{rem}
In the proof of Lemma~\ref{lem_finite} it is considered that $\alpha_k$ is independent of $\alpha_{k+1}$, \textit{i.e.}, the variation rate is arbitrary. If bounded variation rates are to be considered, then one should properly relate the parameters $\alpha_k$ and $\alpha_{k+1}$.
\end{rem}

\section{Numerical Experiments}
\label{nex}

In this section a comparative analysis among the conditions proposed in this paper and the available results from the literature are presented. The routines were implemented in Matlab R2015a, by using the packages YALMIP~\citep{Lof:04},  ROLMIP~\citep{AOP:19} and the solver SeDuMi~\citep{Stu:99}.

\subsection*{Example~1}
Consider the following switched discrete-time LPV system borrowed from~\cite{HDI:06b} with matrices 
\begin{equation*}
\hat{A}_{\sigma}\left(k \right )= A_{0\sigma}+D_{\sigma}F\left(k \right )E_{\sigma}    
\end{equation*}
where $F(k)= \rho(k)$ and $\rho(k) \in\ [-1,1]$
\begin{equation*}
\begin{aligned}
    A_{\sigma1}& = A_{0\sigma}+\rho D_{\sigma}E_{\sigma} \\ A_{\sigma2}& = A_{0\sigma}-\rho D_{\sigma}E_{\sigma}
\end{aligned}
\end{equation*}
with
\begin{equation*}
\begin{aligned}
    A_{01}& =  \begin{bmatrix}
    0.2 & 0.2 & 0.3  & 0.1  & -0.5 \\ 
    0.8 &  0  & -0.1 & -0.3 & 0.3\\ 
    0 & -0.3  & -0.4  & 0   & 0\\ 
    0 & 0.3 & 0.1 & 0.3 & 0.5\\ 
    -0.2 & 0 & 0 & 0 & 0.1 
    \end{bmatrix} \\
    A_{02}& = \begin{bmatrix}
     -0.7 & -0.7 & 0 & 0 & 0.2 \\ 
    0.5 & 0.3 & 0.3 & -0.3 & 0 \\ 
    0.3 & 0.4 & 0.3 & 0.6 & 0.3 \\ 
    0.3 & -0.8 & 0 & 0 & 0 \\ 
    0.1 & -0.7 & 0.1 & -0.3 & 0.3 
    \end{bmatrix} \\
    \sigma& \in [1,2]\\
    D_{1}^{T}& = \begin{bmatrix}
    0.2 & 0.5 & -0.1 & 0.3 & 0.2
    \end{bmatrix} \\
    D_{2}^{T}& = \begin{bmatrix}
    -0.5 & 0.38 & 0.5 & 0.2 & 0.5
    \end{bmatrix} \\
    E_{1}& = \begin{bmatrix}
    -0.3 & -0.3 & -0.5 & 0.2 & 0.3
    \end{bmatrix} \\
    E_{2}& = \begin{bmatrix}
    -0.2 & 0.1 & -0.1 & -0.05 & 0.7
    \end{bmatrix}
\end{aligned}
\end{equation*}
In this way the switched discrete-time LPV system can be written as:
\begin{equation*}
\begin{aligned}
    A_{1}(\alpha_k)& = \alpha_{k,1}A_{11}+\alpha_{k,2}A_{12} \\ A_{2}(\alpha_k)& = \alpha_{k,1}A_{21}+\alpha_{k,2}A_{22}
\end{aligned}
\end{equation*}
For this example, Lemma~\ref{lema3} is able to provide a solution with the use of $30$ scalar decision variables and $210$ LMI rows. On the other hand, the result presented in~\citet{HDI:06} makes use of $160$ scalar decision variables and $160$ LMI rows. It is also important to emphasize that the method proposed in~\cite{XWHX:03} fail to find a solution in this case. 

The Lyapunov function obtained from Lemma~\ref{lema3} is composed by two components $V(x(k))=V_1 + V_2$ with
\begin{align*}
    V_1&=x(k)^T P_1x(k) \\
    V_2& = x(k)^TA(\xi(k),\alpha_k)^TP_2A(\xi(k),\alpha_k)x(k)
\end{align*}

\begin{figure}[ht!]
\begin{psfrags}
\psfrag{60}[r][c]{\scriptsize $60$}
\psfrag{50}[r][c]{}
\psfrag{40}[r][c]{\scriptsize $40$}
\psfrag{30}[r][c]{}
\psfrag{20}[r][c]{\scriptsize $20$}
\psfrag{10}[c][b]{\scriptsize $10$}
\psfrag{0}[r][c]{\scriptsize $0$}
\psfrag{2}[c][b]{\scriptsize $2$}
\psfrag{4}[c][b]{\scriptsize$4$}
\psfrag{6}[c][b]{\scriptsize $6$}
\psfrag{8}[c][b]{\scriptsize$8$}
\psfrag{12}[c][b]{\scriptsize$12$}
\psfrag{14}[c][b]{\scriptsize$14$}
	    \psfrag{ky}[c][c]{$V(x(k))$}
	    \psfrag{kx}[c][c]{$k$}
\epsfxsize=9.9cm
\centerline{\epsffile{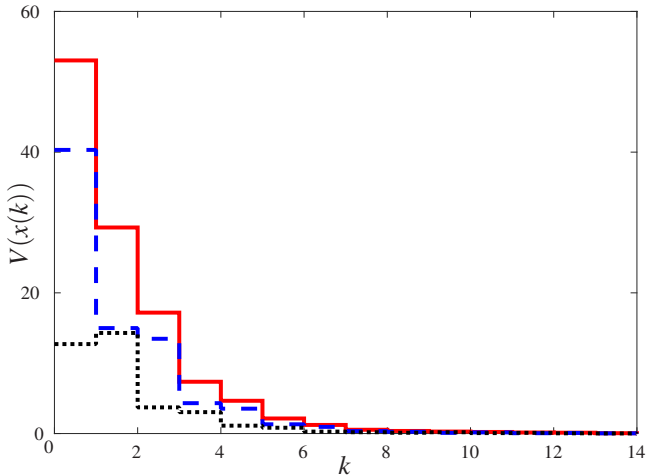}}
\caption{Time evolution of the Lyapunov function $V(x(k))$ (solid {\color{red} red} line) and its components $V_1$ (dashed {\color{blue} blue} line), and $V_2$ (black dotted line), along the trajectories of the LPV discrete-time switched system}
		\label{fig1}
	\end{psfrags}
\end{figure}

\begin{figure}[ht!]
\begin{psfrags}
\psfrag{0.4}[r][c]{\scriptsize $0.4$}
\psfrag{0.5}[r][c]{}
\psfrag{0.6}[r][c]{}
\psfrag{0.7}[r][c]{}
\psfrag{0.8}[r][c]{\scriptsize $0.8$}
\psfrag{0.9}[r][c]{\scriptsize $0.9$}
\psfrag{1}[r][c]{\scriptsize $1$}
\psfrag{0}[c][c]{\scriptsize $0$}
\psfrag{2}[c][c]{\scriptsize $2$}
\psfrag{4}[c][c]{\scriptsize $4$}
\psfrag{6}[c][c]{\scriptsize $6$}
\psfrag{8}[c][c]{\scriptsize $8$}
\psfrag{10}[c][c]{\scriptsize $10$}
\psfrag{12}[c][c]{\scriptsize $12$}
\psfrag{14}[c][c]{\scriptsize $14$}
	    \psfrag{py}[c][c]{$\alpha_{k,1}$}
	    \psfrag{px}[c][c]{$k$}
\epsfxsize=9.9cm
\centerline{\epsffile{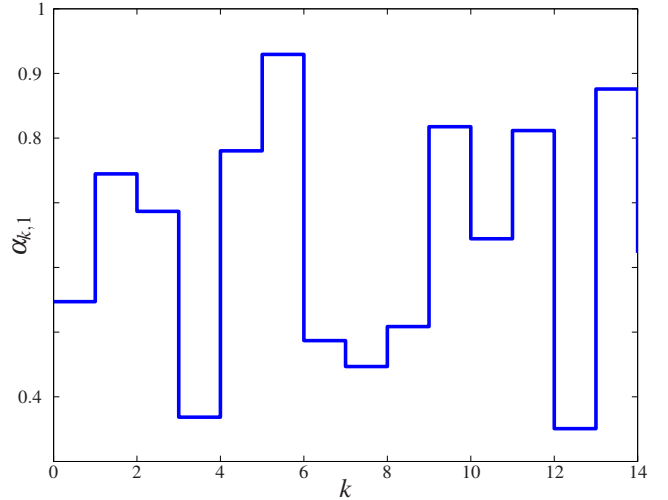}}
\caption{Temporal evolution of the time-varying parameter $\alpha_{k,1}$.}
		\label{fig2}
	\end{psfrags}
\end{figure}

Figure~\ref{fig1} depicts the evolution of the Lyapunov function $V(x(k))$ (solid {\color{red} red} line), $V_1$ (dashed {\color{blue} blue} line), and $V_2$ (black dotted line), along the trajectories of the LPV discrete-time switched system. Note that $V_2$ is not monotonically decreasing along the trajectories. It is important to remember that the switched system is subjected to the action of the time-varying parameters. Figure~\ref{fig2} shows the behavior of the time-varying parameter $\alpha_{k,1}$ over time. The switching rule may be arbitrary, but in this case it has been considered to change each iteration, starting in mode~$1$. 

\subsection*{Example~2}

This example is adapted from~\cite{LD:06}. Consider the switched discrete-time LPV system 
\begin{equation*}
    A_{1}(\alpha)=\begin{bmatrix}
\beta & \beta \\ 
0 & 0
\end{bmatrix}, \quad
A_{2}(\alpha)=\begin{bmatrix}
-\beta & 0 \\ 
\beta & -\beta
\end{bmatrix}        
\end{equation*}
where $\beta$ is the time-varying parameter $\beta \in [-\theta ,\theta]$. The main goal is to find the maximum value of $\theta$ such that it is possible to certify the stability of the system. For this end, Theorem~\ref{teo1} and Corollary~\ref{cor1} will be employed with different values of~$N$. Table~\ref{tab2} presents the maximum values of $\theta$, as well as the number of scalar decision variables $N_V$ and LMI rows $N_R$ obtained for each method and different values of~$N$.
\begin{table}[ht]
\begin{center}
\caption{Maximum values for $\theta$, number of scalar decision variables $N_V$, and number of LMI rows $N_R$  when considering different values of $N$ in Theorem~\ref{teo1} and in Corollary~\ref{cor1}.}
\label{tab2}
\begin{tabular}{ccccc}
 \\\hline
\multicolumn{5}{c}{Theorem~\ref{teo1}}\\
$N$  & $2$ & $3$ & $4$ & $5$  \\ \hline
$\theta_{Max}$ & $0.7413$ & $0.7430$ & $0.7430$ & $0.7430$  \\
$N_{V}$ & $6$ & $9$ & $12$ & $15$ \\ 
$N_{R}$ & $84$ & $504$ & $3024$ & $18144$ \\ \hline 
\multicolumn{5}{c}{Corollary~\ref{cor1}}\\
$N$  & $2$ & $3$ & $4$ & $5$  \\ \hline
$\theta_{Max}$ & $0.7547$ & $0.7714$ & $0.7723$ & $0.7723$  \\
$N_{V}$ & $12$ & $18$ & $24$ & $30$ \\ 
$N_{R}$ & $144$ & $864$ & $5184$ & $31104$ \\\hline 
\end{tabular}
\end{center}
\end{table}

It can be seen that higher values of~$N$ provide less conservative results, notably when using Corollary~\ref{cor1}. However, the best results come with a greater computational burden. The technique~\citep[Theorem~3]{HDI:06} is also applied to the current example, resulting in $N_V = 28$, $N_R = 64$ and ${\theta_{Max} = 0.7548}$. In this sense, the method presented in Corollary~\ref{cor1} is able to assess the stability with a broader interval for the uncertainty $\beta$ with a smaller number of scalar decision variables.

\section{Conclusions}
\label{con}

New stability conditions for discrete-time LPV switched systems have been proposed in this paper. The system is supposed to be affected by arbitrary switching, where each mode depends on time-varying parameters lying within a polytopic domain. The proposed conditions stem from the application of Lyapunov functions depending not only on the current states, but also on shifted states. Numerical experiments illustrate the advantages of the proposed method, which is capable of certifying the stability of LPV switched systems by using less variables than other techniques from the literature. Additionally, the proposed Lyapunov function may depend on an arbitrary number of shifted states, and increasing such number leads to less conservative conditions, as shown in the experiments. As future research the authors are investigating the stabilization problem for LPV switched systems.



\end{document}